\documentclass[
 reprint,
 amsmath,
 amssymb,
 aps,
 superscriptaddress,
]{revtex4-2}

\usepackage{graphicx}
\usepackage{dcolumn}
\usepackage{bm}
\usepackage{siunitx}
\DeclareSIUnit\torr{Torr}
\DeclareSIUnit\bar{bar}
\usepackage{hyperref}

\begin{document}

\preprint{APS/123-QED}

\title{Superconducting Parallel-Plate Resonators for the Detection of Single Electron Spins}

\author{André Pscherer}
\affiliation{Quantronics Group, Service de Physique de l’État Condensé (CNRS, UMR 3680), \\
IRAMIS, CEA-Saclay, Université Paris-Saclay, 91191 Gif-sur-Yvette, France}
\author{Jannes Liersch}
\affiliation{Quantronics Group, Service de Physique de l’État Condensé (CNRS, UMR 3680), \\
IRAMIS, CEA-Saclay, Université Paris-Saclay, 91191 Gif-sur-Yvette, France}
\author{Patrick Abgrall}
\affiliation{Quantronics Group, Service de Physique de l’État Condensé (CNRS, UMR 3680), \\
IRAMIS, CEA-Saclay, Université Paris-Saclay, 91191 Gif-sur-Yvette, France}
\author{Andrew D. Beyer}
\affiliation{Jet Propulsion Laboratory, California Institute of Technology, Pasadena, California 91109, USA}
\author{Fabien Defrance}
\altaffiliation[Current address: ]{Catalan Institute of Nanoscience and Nanotechnology (ICN2), CSIC and BIST, Campus UAB, Bellaterra 08193, Barcelona, Spain}
\affiliation{Jet Propulsion Laboratory, California Institute of Technology, Pasadena, California 91109, USA}
\author{Sunil R. Gowala}
\affiliation{California Institute of Technology, Pasadena, California 91125, USA}
\author{Hélène Le Sueur}
\affiliation{Quantronics Group, Service de Physique de l’État Condensé (CNRS, UMR 3680), \\
IRAMIS, CEA-Saclay, Université Paris-Saclay, 91191 Gif-sur-Yvette, France}
\author{James O'Sullivan}
\affiliation{Quantronics Group, Service de Physique de l’État Condensé (CNRS, UMR 3680), \\
IRAMIS, CEA-Saclay, Université Paris-Saclay, 91191 Gif-sur-Yvette, France}
\author{Emmanuel Flurin}
\affiliation{Quantronics Group, Service de Physique de l’État Condensé (CNRS, UMR 3680), \\
IRAMIS, CEA-Saclay, Université Paris-Saclay, 91191 Gif-sur-Yvette, France}
\author{Patrice Bertet}
\affiliation{Quantronics Group, Service de Physique de l’État Condensé (CNRS, UMR 3680), \\
IRAMIS, CEA-Saclay, Université Paris-Saclay, 91191 Gif-sur-Yvette, France}
\email{patrice.bertet@cea.fr}


\date{\today}

\begin{abstract}
We introduce a multilayer superconducting microwave resonator with sub-Ohm impedance optimized for high coupling strength to single electron spins. The design minimizes the magnetic far-field and therefore achieves a Purcell factor $F_P > 10^{15}$. We show several ways to fabricate this type of resonator and present resonators with an intrinsic $Q$-factor exceeding \num{2e+4} at the single-photon level. We further characterize these resonators in magnetic fields up to \SI{500}{\milli\tesla}. Finally, we evaluate the impact of the achievable Purcell factor on single-spin detection through photon counting and dispersive readout.
\end{abstract}

\maketitle


\section{Introduction}

Single electron spins have proven their applicability for quantum sensing \cite{Lovchinsky2016,Abobeih2018,Uysal2023,Du2024,VandeStolpe2024,Sellies2025}, quantum information processing \cite{Taminiau2014,Awschalom2018,Sun2018,Chen2020,Wan2020,Bayliss2020,Ruskuc2022,Abobeih2022} and quantum communication \cite{Pfaff2013,Dibos2018,Bhaskar2020,Stevenson2022,Stas2022,Ourari2023,Coste2023,Hogg2025} in the recent years. The main readout method, however, using a spin-dependent optical transition, limits the range of available systems. On the other hand, magnetic dipole coupling to the magnetic component of the microwave field is universal among all electron spin systems. This coupling is very small in free space, but can be significantly increased using the Purcell effect \cite{Purcell1946,Bienfait2016,Ranjan2020b,Wang2023}: At resonance, a spin emits into a resonator at a rate $\Gamma_\mathrm{P} = 4 {g_0}^2 / \kappa$, where $g_0$ is the spin-resonator coupling strength and $\kappa$ is the resonator decay rate. In recent experiments, a radiative spin lifetime of $\Gamma_\mathrm{P}^{-1} = \SI{0.8}{\milli\second}$ was shown for a single Er$^{3+}$ ion in CaWO$_4$ \cite{Travesedo2025, OSullivan2025}. Compared to the free-space radiative decay rate $\gamma$, this is a speedup by a Purcell factor of $F_P = \Gamma_\mathrm{P} / \gamma = 10^{14}$.

The Purcell factor $F_P$ achieved in these experiments is sufficient for the detection of single Er$^{3+}$ ions in CaWO$_4$, which have an effective Landé-factor of $g = \num{8.38}$. Most spins however, have a Landé-factor close to that of a free electron spin $g \approx 2$ and therefore an emission rate $\Gamma_\mathrm{P}$ which is 17 times lower. Despite interest in the spectroscopy of these spins \cite{Keyser2020,Gimeno2020}, the low emission rates have prevented single-spin spectroscopy so far. Also for quantum computation applications \cite{Kane1998}, larger coupling strengths $g_0$ would be desirable, since the speed of resonator-mediated 2-spin-qubit gates scales as $\sim {g_0}^2$ \cite{Blais2021}.

To achieve a high coupling strength $g_0$, the resonator must concentrate the magnetic component of its mode in a small volume. In the experiments \cite{Bienfait2016,Ranjan2020b,Wang2023,Travesedo2025,OSullivan2025} this is achieved using a planar superconducting resonator consisting of an interdigitated finger capacitor shunted by a nanowire. The current passing through the nanowire induces a magnetic field, which couples spins within a few hundred nanometers to the resonator. However, parasitic inductances create magnetic fields in other regions around the resonator, for example around the capacitors.

In this work, we present resonators which largely reduce the parasitic inductance, thereby increasing the coupling strength $g_0$ between the resonator and a single spin by a factor of 5. With all other experimental parameters being equal, this increases the radiative emission rate by a factor of 25.
In section~\ref{sec:simulation} we explain the geometry and field distribution of the parallel-plate resonators as well as different ways to connect them to a coaxial transmission line. Section~\ref{sec:fabrication} shows three different methods to fabricate the resonators, which are characterized in section~\ref{sec:characterization} in terms of their losses and susceptibility to magnetic fields. In section~\ref{sec:outlook} we will elaborate how the resonators improve existing spin detection protocols and which new ones are enabled, before concluding in section~\ref{sec:conclusion}.

\section{Simulation} \label{sec:simulation}

\subsection{Working principle}

\begin{figure*}
\includegraphics{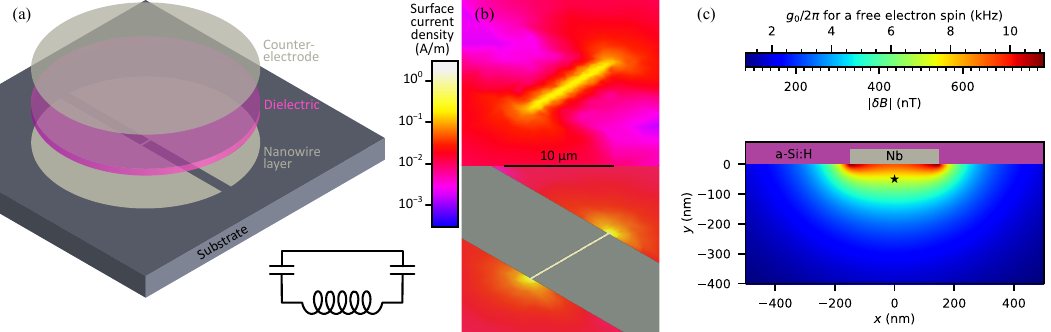}
\caption{(a) Exploded view of the resonator (not to scale) and equivalent lumped-element circuit. (b) Current density in the nanowire region (bottom) and the counter-electrode. (c) Cross-section through the nanowire showing the magnetic field ZPF $|\delta B|$ and the coupling strength between the resonator and a free electron spin $g_0$.}  \label{fig:resonator_layout}
\end{figure*}

The resonator consists of three layers, as shown in Fig.~\ref{fig:resonator_layout}a, which we will refer to as the superconducting \textit{nanowire layer}, the \textit{dielectric} and the superconducting \textit{counter-electrode}. The semi-disk areas on both sides of the nanowire form capacitors with the counter-electrode. The resonator mode of interest is the one in which the charge oscillates between the two semi-disks, creating a high current density in the nanowire. For this mode, the resonator is well approximated by the lumped-element $LC$-circuit shown as inset.

The current in the nanowire layer induces a mirror current in the counter-electrode, of equal amplitude and opposite phase. The current density is concentrated in the counter-electrode at the position the nanowire (see Fig.~\ref{fig:resonator_layout}b), creating a virtual image of the nanowire. Hence, the magnetic field is confined to the volume between the two superconducting layers, and canceled outside. Note, however, that, due to the larger current density in the nanowire electrode, the near-field around the nanowire is dominated by the nanowire current and not by its mirror image. This area is where the spins of interest must be located. The coupling rate is
\begin{equation}
    g_0 = \frac{1}{2} \left| {\delta \vec{B}}_\perp (\vec{r}) \, \underline{\underline{\gamma}} \right|,
\end{equation}
where ${\delta \vec{B}}_\perp (\vec{r})$ are the zero-point fluctuations (ZPF) of the magnetic field perpendicular to the spin quantization axis and $\underline{\underline{\gamma}}$ is the gyromagnetic tensor of the spin. Given that the current ZPF in the nanowire $\delta I$ are proportional to the magnetic field ZPF, we use it as a quantifiable measure for the coupling between the resonator and spins, without requiring the exact spin location or its gyromagnetic tensor. Without loss of generality, the plots in Fig.~\ref{fig:resonator_layout}b and c quantify the current ZPF for a resonator with the following parameters: Capacitor diameter \SI{825}{\micro\meter}, nanowire length $l = \SI{10}{\micro\meter}$, nanowire width $w = \SI{300}{\nano\meter}$, superconducting film thickness $t = \SI{50}{\nano\meter}$, dielectric material thickness \SI{500}{\nano\meter} with $\varepsilon_r = 11.9$, resulting in a resonance frequency of $f_r = \SI{7.5}{\giga\hertz}$ and current ZPF of $\delta I = \SI[round-mode=places,round-precision=0]{394.910691674662}{\nano\ampere}$. More details about these Ansys Electromagnetics Suite simulations can be found in the Supplementary Information \cite{SupMat}.
Fig.~\ref{fig:resonator_layout}c shows the magnetic field ZPF in the substrate below the nanowire. A spin \SI{50}{\nano\meter} below the center of the wire, as indicated by the star, experiences ZPF of ${| \delta \vec{B}}_\perp | = \SI[round-mode=places,round-precision=0]{511.49829289974457}{\nano\tesla}$, which gives rise to a resonator-spin coupling of $g_0 / 2\pi = \SI[round-mode=places,round-precision=1]{7.167357406038283}{\kilo\hertz}$ for a free electron spin. Er$^{3+}$ in CaWO$_4$, a commonly used spin species with more than 4 times higher gyromagnetic ratio, even reaches $g_0 / 2\pi = \SI[round-mode=places,round-precision=1]{30.03122753130041}{\kilo\hertz}$ at this position. For comparison with other work on low-impedance resonators, we specify that the impedance of this mode is $Z = \frac{\hbar}{2} \left( \frac{\omega_r}{\delta I} \right)^2 = \SI[round-mode=places,round-precision=2]{0.750811638157477}{\ohm}$ and its inductance is $L = \frac{\hbar \omega_r}{2 {\delta I}^2} = \SI[round-mode=places,round-precision=1]{15.9327178058248}{\pico\henry}$. A significant contribution to the latter is the kinetic inductance of the nanowire $L_k = L_{k, \Box} \frac{l}{w} = \SI[round-mode=places,round-precision=1]{6.66666}{\pico\henry}$, where $L_{k, \Box} = \SI{0.2}{\pico\henry/\Box}$ is the kinetic inductance per square for a $t = \SI{50}{\nano\meter}$ thick film of niobium. To achieve a smaller inductance, the nanowire can be shorter, however, this would decrease the detection volume.

Another commonly used metric for the coupling of Purcell resonators is the mode volume $V$. For a spin coupling to the magnetic field, it is defined as \cite{Choi2023}
\begin{equation}
    V = \frac{\iiint \frac{| \vec{B} (\vec{r}) |^2}{2 \mu_0} \, d^3\vec{r}}{\max_{\vec{r}} \{ \frac{| \vec{B} (\vec{r}) |^2}{2 \mu_0} \} },
\end{equation}
that is, the ratio of the integrated magnetic energy of the resonator mode over the peak energy density. However, since a significant fraction of the inductive energy is the kinetic energy of the Cooper pairs, we replace the magnetic field energy by the overall inductive energy in the mode. With the resonator in the vacuum state, this is half the zero-point energy, $\frac{1}{4} h f_r$. Furthermore, we replace the magnetic field amplitude $| \vec{B} (\vec{r}) |$ by the amplitude of its vacuum fluctuations $| \delta \vec{B}_\perp (\vec{r}) |$. The mode volume modified in this way is
\begin{equation}
   V^* = \frac{\frac{1}{4} h f_r}{\max_{\vec{r}} \{ \frac{ |\delta \vec{B}_\perp (\vec{r}) |^2}{2 \mu_0} \}} = \SI[round-mode=places,round-precision=2]{4.950375463365}{\micro\meter^3} = \num[round-mode=places,round-precision=1]{7.75103719282791E-14} \lambda^3,
\end{equation}
where $\lambda = c / f_r$ is the wavelength of the $f_r = \SI{7.5}{\giga\hertz}$ microwave tone. This is more than a factor of 20 smaller than the mode volume in previously used resonators with a nanowire measuring $\SI{300}{\nano\meter} \times \SI{50}{\nano\meter}$ in cross-section \cite{Travesedo2025,OSullivan2025}. The mode volume of \SI{10}{\micro\meter^3} in Ref.~\cite{Wang2023} was estimated with a different convention.

The Purcell factor is
\begin{equation}
    F_P (\vec{r}) = \frac{3}{4 \pi^2} \frac{Q}{V^* / \lambda^3} \left| f(\vec{r}) \right|^2,
\end{equation}
where $|f (\vec{r})| = |\delta \vec{B}_\perp (\vec{r}) | / \max_{\vec{r}} \{ |\delta \vec{B}_\perp (\vec{r}) | \}$ is a normalized function describing the spatial profile of the resonator mode. For a realistic $Q$-factor of $Q = 10^4$, the Purcell factor can be as high as $F_P = \num[round-mode=places,round-precision=1]{9.80396375882033e+15}$ for a spin at the field maximum $|f (\vec{r})| = 1$. This would be just at the corners of the nanowire. However, spins close to the surface are known to have short coherence times \cite{Sangtawesin2019,Ranjan2021}. Therefore, we turn to the example spin \SI{50}{\nano\meter} below the center of the nanowire, which experiences a Purcell factor of $F_P = \num[round-mode=places,round-precision=1]{4.06659293696443E15}$.

While the sequence of the layers shown in Fig.~\ref{fig:resonator_layout}a is designed to couple to spins in the substrate, one can reverse their order, such that the nanowire layer is on top. Resonators of this kind can be used for sensing and spectroscopy experiments with the sample placed on top of the nanowire \cite{Keyser2020,Gimeno2020}.

\subsection{Coupling to the transmission line} \label{sec:coupling_transmission_line}

\begin{figure*}
\includegraphics{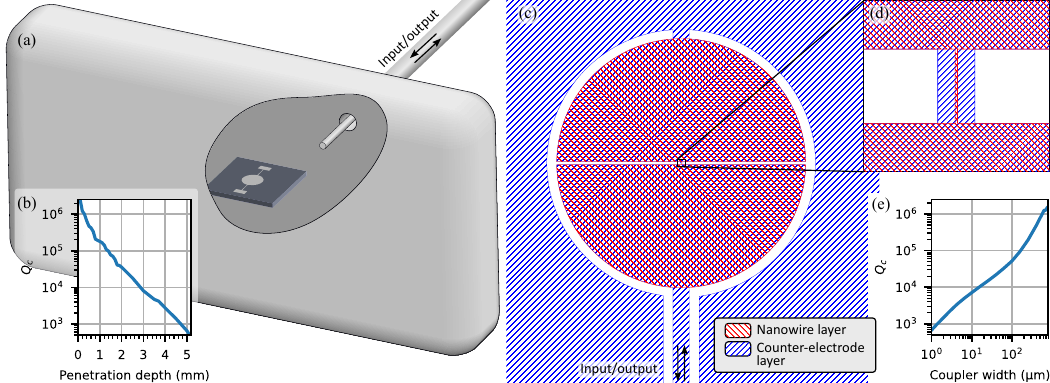}
\caption{(a) A parallel-plate resonator couples via an antenna to a 3D box mode, which couples to a coaxial cable via the cable pin. (b) Coupling $Q$-factor of the parallel-plate resonator to the transmission line as a function of the cable pin penetration depth. (c) Superconducting layers of a waveguide-coupled parallel-plate resonator with a close-up (d) of the nanowire region. (e) The coupling to the coplanar waveguide is controlled through the width of the constriction in the counter-electrode layer.}
\label{fig:coupling_transmission_line}
\end{figure*}

\subsubsection{3D cavity}

A common method to couple superconducting resonators and qubits to a transmission line is via the mode of a 3D metal cavity \cite{Paik2011}. The electric dipole oscillation of the resonator couples to a 3D box mode, which is then picked up by the center conductor of an SMA connector. For the parallel-plate resonator, electric dipole oscillations in the nanowire layer are compensated by the opposite-sign oscillations in the counter-electrode, diminishing the coupling to a homogeneous electric field mode. However, if an antenna with pads at the ends is added to the nanowire layer (see Fig.~\ref{fig:coupling_transmission_line}a), the coupling to the box mode can be recovered. The end pads form coupling capacitors with the side walls of the 3D cavity. Hence, the coupling is sensitive to the distance from the walls. The 3D box has dimensions of $\SI{6}{\milli\meter} \times \SI{17.9}{\milli\meter} \times \SI{34}{\milli\meter}$ and the antenna is \SI{4}{\milli\meter} along the shortest box dimension, with pads measuring $\SI{400}{\micro\meter} \times \SI{400}{\micro\meter}$ (sample~3) or $\SI{1200}{\micro\meter} \times \SI{400}{\micro\meter}$ (samples~4~\&~5). This results in a coupling between the resonator and box mode of $\sim \SI{100}{\mega\hertz}$. Using the 3D cavity as a Purcell filter to the transmission line allows us to control the coupling of the parallel-plate resonator to the transmission line by tuning the linewidth of the 3D cavity. The linewidth increases as the center conductor of the SMA connector penetrates further into the 3D cavity. The coupling rate can be tuned by 3 orders of magnitude in this way (see Fig.~\ref{fig:coupling_transmission_line}b). Further linewidth tuning can be achieved by using 3D cavities with different fundamental mode frequencies.

\subsubsection{Coplanar waveguide}

An alternative way of coupling the resonator to a transmission line is via a coplanar waveguide in the counter-electrode layer. In this case, the counter-electrode is connected to the center conductor of a coplanar waveguide, as shown in Fig.~\ref{fig:coupling_transmission_line}c,d. A constriction in the counter-electrode near the nanowire couples the resonator inductively to the transmission line. The width of the constriction determines the coupling rate. Fig.~\ref{fig:coupling_transmission_line}e shows that the coupling rate can be tuned by 3 orders of magnitude. An advantage of this coupling method is that it is possible to send DC or radio-frequency current through the counter-electrode constriction, which could be used to dynamically create a magnetic field offset to detune nearby electron spins from the resonator or drive nuclear spin transitions \cite{Wen2025}.

Lastly, we mention the possibility of galvanically coupling the resonators to a coplanar waveguide. For many types of resonators, this would lead to a low coupling $Q$ factor. However, parallel-plate resonators have a very low impedance, such that the impedance mismatch between the resonator and a \SI{50}{\ohm} transmission line leaves the resonator with a moderate $Q_c = \frac{Z}{\SI{50}{\ohm}} = \num[round-mode=places,round-precision=0]{66.594599043113}$. We found this coupling method particularly useful in the early days of the project to achieve a broadband coupling to resonators. In this case, we wirebonded the antenna-pads to a coplanar waveguide. A microscope image of this is found in the Supplementary Information \cite{SupMat}.

\section{Fabrication} \label{sec:fabrication}

\subsection{Architectures}

Three different routes to fabricate parallel-plate resonators were explored: Additive fabrication on a given substrate, fabrication on a silicon membrane, and subtractive fabrication starting from a Nb-Si-Nb trilayer deposited on a substrate.

\subsubsection{Additive fabrication}

The additive fabrication is suitable for coupling to spins in the substrate, for example, Er$^{3+}$-doped CaWO$_4$. Hence, the nanowire layer is on the bottom. It is fabricated by sputtering Nb on the substrate, which is then patterned by reactive ion etching (RIE) using an Al hard mask created by a combination of optical and electron-beam lift-off, as detailed in Sec.~\ref{sec:processes}. Next, the dielectric layer is deposited. In this work, we use amorphous hydrogenated silicon (a-Si:H) deposited at the Jet Propulsion Laboratory (JPL) Microdevice Laboratory clean room. The a-Si:H recipe is detailed in section~\ref{sec:a-Si:H}.  With demonstrated zero-temperature single-photon loss tangents, $\tan \delta$ of a few $10^{-6}$  ($Q_i > 10^5$), the a-Si:H thin films developed at JPL and Caltech \cite{Defrance2024} are among the lowest-loss amorphous dielectrics to date. Lastly, the counter-electrode layer is created by sputtering Nb and patterning it with RIE using only optical lithography, because there are no small features in this layer.

\subsubsection{Silicon membrane}

It has been shown that capacitors with monocrystalline silicon as dielectric may have low intrinsic losses, with internal $Q$-factors up to $Q_i = \num{2e+5}$ in the single-photon regime \cite{Weber2011,McFadden2025}. Monocrystalline silicon is used in the form of membranes, as in Ref.~\cite{Weber2011}, purchased from Silson Ltd. in the form of $4 \times 4$ membrane multi-frame arrays. Each frame is processed individually, though in principle the whole array can be processed in parallel. Each frame is \qtyproduct{5 x 5}{\milli\meter} with the device layer free-standing in the central \qtyproduct{2.5 x 2.5}{\milli\meter}. The device layer is a \SI{2}{\micro\meter} thick float-zone high-resistivity silicon and the frame (the leftover of the handle wafer) is a \SI{300}{\micro\meter} thick silicon wafer. The nanowire layer is fabricated in the same way as in the additive fabrication, on top of the membrane. The counter-electrode is not lithographically patterned, but by sputtering the Nb through a coarsely aligned Kapton foil mask. The lack of control of the exact size and shape of the counter-electrode layer does not pose a problem to the resonator performance, because the area of the parallel-plate capacitors is given by the overlap of the counter-electrode and nanowire layer, with the latter being precisely patterned. We note that this also facilitates increasing the resonance frequency of resonators in post-fabrication. To reduce the size of the capacitor, we only have to etch one of the parallel plates.

With silicon membranes we only tested antenna-to-3D-box-coupled resonators. They can be coupled to spins by either bonding them to a spin-doped substrate \cite{Sigillito2017,Dibos2018,Wen2025}, doping the spins directly into the Si membrane \cite{Weiss2021} or placing spins on the nanowire, for example using a micromanipulator \cite{Keyser2020}, nano-printing techniques \cite{Hail2019,Eichhorn2025} or simple spin-coating or drop-casting.

\subsubsection{Trilayer etching}

It is known that a significant fraction of intrinsic losses of superconducting resonators originates from the interfaces between layers \cite{McRae2020}. We therefore investigated a fabrication approach, in which one starts with a Nb-Si-Nb trilayer on the substrate. Ideally, this trilayer can be fabricated without breaking the process vacuum, or at least with only a few minutes of atmospheric exposure between the steps. However, to benefit from JPL a-Si:H thin films, we have decided to split the fabrication steps between JPL and CEA: As for the additive fabriaction, the Nb layers were deposited in the cleanroom of CEA~Saclay, whereas the a-Si:H layer was deposited in the JPL cleanroom. The dielectric losses are therefore expected to be comparable for both devices. We nevertheless would like to share the fabrication procedure and characterization results (Sec.~\ref{sec:characterization}) of these samples.

Restricting ourselves to etching methods which do not significantly under-etch, the resulting area of the top Nb layer must be a sub-set of the area of the bottom Nb layer. Hence, the top layer must contain the nanowire. Furthermore, since the 3D-box-coupling requires an antenna in the nanowire layer without correspondance in the counter-electrode layer, we decided to implement the waveguide coupling.

\begin{figure}
\includegraphics{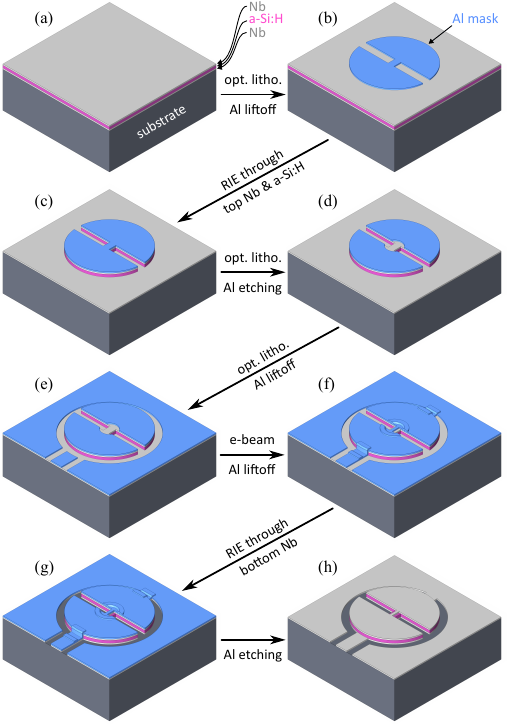}
\caption{Fabrication of a waveguide-coupled parallel-plate resonator from a Nb-Si-Nb trilayer. See the text for an explanation of the steps. The features of the resonator are not to scale.}
\label{fig:trilayer_fabrication}
\end{figure}

The fabrication steps are displayed in Fig.~\ref{fig:trilayer_fabrication}. First, we optically pattern an aluminum hard mask covering the nanowire layer. Instead of a constriction corresponding to the nanowire, the mask has a constriction corresponding to the resulting width of the coupler in the counter-electrode layer (see Fig.~\ref{fig:trilayer_fabrication}b). Second, in the unprotected area, the top Nb and the a-Si:H layers are reactive-ion etched away (Fig.~\ref{fig:trilayer_fabrication}c). Third, a small area of the aluminum mask around the constriction is removed using optically patterned resist and MF-319 (Fig.~\ref{fig:trilayer_fabrication}d). Fourth, more parts of the aluminum mask are created to define the coarse features of the waveguide and the groundplane, using optical-lithography Al liftoff (Fig.~\ref{fig:trilayer_fabrication}e). Fifth, using electron-beam lithography Al liftoff, we define the nanowire and the connections of the resonator counter-electrode to the waveguide and groundplane (Fig.~\ref{fig:trilayer_fabrication}f). This completes the Al mask for the last step of reactive ion etching (Fig.~\ref{fig:trilayer_fabrication}g), in which the bottom Nb layer is patterned, as well as the nanowire in the top Nb layer. Finally, the Al mask is removed (Fig.~\ref{fig:trilayer_fabrication}h).

\subsection{Processes} \label{sec:processes}

\subsubsection{Nb deposition}

To remove oxides from the silicon substrate, the substrate is bathed in \SI{5}{\percent} HF for \SI{2}{\minute}, followed by \SI{1}{\minute} in deionized water, and transferred into the Alcatel MODEL-NO sputtering machine and start evacuating the process chamber within less than \SI{5}{\minute}. A liquid nitrogen trap allows base pressures around \SI{1e-7}{\milli\bar} to be reached. After cleaning the sputtering target with \SI{12.1}{\micro\bar} of argon plasma at a voltage of \SI{292}{\volt} for \SI{5}{\minute}, we open the shutter and deposit Nb at a rate of \SI{1}{\nano\meter\per\second}.

\subsubsection{a-Si:H deposition} \label{sec:a-Si:H}

To remove oxides from the Nb layer, the samples are cleaned in a 10:1 buffered oxide etch for \SI{3}{\minute}, followed by a deionized water rinse and blown dry in N$_2$. They are then transferred into an Oxford Plasmalab System 100 ICP 380 at JPL MDL for ICP-PECVD. \SI{465}{\nano\meter} of a-Si:H were deposited on samples~1-4 at a temperature of \SI{350}{\celsius} in \SI{10}{\milli\torr} of SiH$_4$. \SI{490}{\nano\meter} of a-Si:H were deposited on sample~5 at a temperature of \SI{300}{\celsius} in \SI{10}{\milli\torr} of SiH$_4$.

\subsubsection{Al hard mask}

We pattern an Al hard mask by a combination of optical and electron-beam patterning of lift-off resist. For coarse structures, \SI{200}{\nano\meter} of LOL2000 is spin-coated and \SI{400}{\nano\meter} of S1805 on the substrate, expose it with a Heidelberg Instruments µMLA and develop the resist for \SI{45}{\second} in MF-319. \SI{35}{\nano\meter} of Al is evaporated at a rate of \SI{1}{\nano\meter\per\second} in a Plassys MEB550s electron beam evaporator. The Al on the resist is lifted off in a \SI{30}{\minute} bath in Microposit Remover 1165 at \SI{80}{\celsius}.

Small features, such as the nanowire, are patterned with an electron beam. The chemicals for this process are chosen to not damage Al hard masks from previous steps, such that an optically patterned Al mask can be complemented with a electron beam Al mask. \SI{400}{\nano\meter} of p(MAA-MMA)-EL10 is spin-coated and \SI{100}{\nano\meter} of PMMA \SI{950}{\kilo\dalton} on the sample and expose it with a \SI{25}{\kilo\volt} beam in a Raith eLINE. The resist is developed in a 3:1~mixture of isopropanol and methyl isobutyl ketone for \SI{45}{\second} and pure isopropanol for \SI{15}{\second}. The Al deposition and liftoff is done in the same way as with the optical lithography.

\subsubsection{Reactive ion etching}

The RIE is done in a Plassys MG200, using a 2:1 mixture of CF$_4$ and Ar at a pressure of \SI{50}{\micro\bar} and a plasma voltage of \SI{208}{\volt}. To monitor the etching progress, the reflection of a \SI{670}{\nano\meter} laser beam is measured during the process. The completion of each Nb layer etch is identified by a steep drop in the reflectivity. While etching through the a-Si:H we see interference fringes, which abruptly stop when the Nb bottom layer is reached.

\section{Resonator characterization} \label{sec:characterization}

The resonators are characterized in dilution refrigerators at temperatures below \SI{20}{\milli\kelvin} or in an adiabatic demagnetization refrigerator at \SI{140}{\milli\kelvin} or \SI{200}{\milli\kelvin}. The setups are described in detail in the Supplementary Information \cite{SupMat}. The reflection coefficient $S_{11}(f)$ of the resonators is measured as a function of the excitation frequency $f$. The spectra are then de-embedded (see Supplementary Information \cite{SupMat}) and fit them, using the ABCD~RF~Fit package \cite{ABCD-RF_Fit}, to extract the intrinsic quality factor $Q_i$ and coupling quality factor $Q_c$. At the single-photon level, $Q_i$ can be strongly reduced due to two-level systems in the material \cite{DeLeon2021}. This is relevant because the Purcell effect depends on the quality factor of the resonator in the absence of photons. Hence, we characterize the resonators at a range of microwave powers in Section~\ref{sec:material_losses}. Furthermore, the magnetic field necessary to tune the Zeeman transition of the spins in resonance with the parallel-plate resonators is on the order of a few hundred \unit{\milli\tesla}. This can shift the resonance frequency and increase the intrinsic losses of resonators. In Section~\ref{sec:magnet}, we characterize the resonators in magnetic fields up to \SI{500}{\milli\tesla}. Table~\ref{tab:samples} provides an overview of the characterized samples.

\begin{table*}
    \caption{Overview of the characterized samples. Microscope images can be found in the supplementary information \cite{SupMat}.}
    \label{tab:samples}
    \begin{ruledtabular}
    \begin{tabular}{lllll}
    Sample number & Coupling & Dielectric & Nanowire width $w$ & Remarks \\ \hline
    1 & waveguide & a-Si:H & \SI{300}{\nano\meter} & fabricated from trilayer \\
    2 & waveguide & a-Si:H & \SI{4000}{\nano\meter} & fabricated from trilayer \\
    3 & antenna & a-Si:H & \SI{4000}{\nano\meter} & fabricated layer by layer \\
    4 & antenna & crystalline Si & \SI{4000}{\nano\meter} & silicon membrane \\
    5 & antenna & a-Si:H & \SI{300}{\nano\meter} & fabricated layer by layer on CaWO$_4$
    \end{tabular}
    \end{ruledtabular}
\end{table*}

\subsection{Material losses} \label{sec:material_losses}

\begin{figure}
\includegraphics{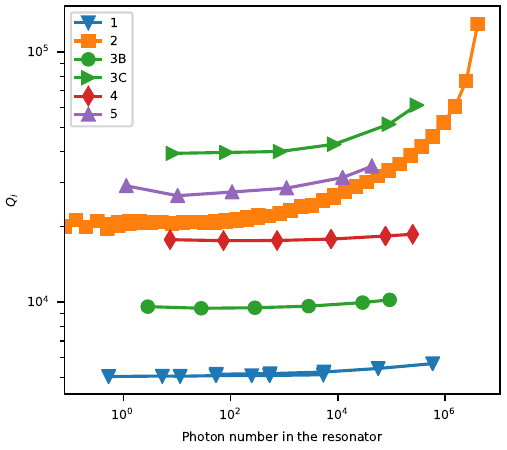}
\caption{Dependence of $Q_i$ of the resonators listed in Table~\ref{tab:samples} with the photon number $\bar{n}$ in the resonator. At low power, two-level systems cause additional losses.} \label{fig:QvsP}
\end{figure}

Fig.~\ref{fig:QvsP} shows the extracted intrinsic quality factors $Q_i$. Example spectra for sample~3 together with an explanation of the de-embedding can be found in the Supplementary Information \cite{SupMat}. Most resonators are found to show internal $Q$-factors above $10^4$ at the single-photon level. However, this is still an order of magnitude lower than the best reported $Q_i$-factors for a-Si:H \cite{Defrance2024} and crystalline silicon \cite{Weber2011,McFadden2025}. This and the fact that resonators on the same chip (e.g. 3B and 3C) can strongly differ in $Q_i$ is an indication that the losses are not purely due to absorption in the dielectric, but have significant contributions from irreproducible defects, which might be suppressed with an optimized fabrication process.

It is also notable that the resonator fabricated with the silicon membrane as dielectric (sample~4) has a $Q_i$ similar to those with a-Si:H dielectric layer. We suspect that this might be in part due to some remaining silicon oxides, as pointed out in Ref.~\cite{Weber2011}. Thermal oxides were removed with a \SI{2}{\minute} bath
 in \SI{5}{\percent} HF directly before the Nb deposition. However, if some of the buried oxide on the bottom side of the membrane remained after the membrane was etched free, this might not be sufficient to fully remove it. A longer de-oxidation step is expected to increase $Q_i$.

Among the characterized resonators, some have a $Q_i$ well beyond \num{2e+4} at the single-photon level, which is sufficient for a resonator with an overall $Q = \left( 1/Q_c + 1/Q_i \right)^{-1}$ on the order of $10^4$. This is the $Q$-factor we use in the quantitative examples in Sec.~\ref{sec:outlook}. For a given $Q_i$, the highest photon extraction rate is achieved when $Q_c = Q_i$, i.e. at critical coupling. For the spectroscopy of single spins, a lower $Q_c$ can be beneficial. The spin linewidth can be up to several \unit{\mega\hertz} due to interaction with the environment \cite{Childress2006,Abe2010}. To Purcell-enhance the whole spin emission spectrum, the resonator linewidth $\kappa$ should be larger than or comparable to the spin linewidth.

\subsection{Magnetic field sweep} \label{sec:magnet}

The magnetic field for the Zeeman splitting has two effects on the superconducting resonators. Firstly, it lowers the Cooper-pair density, leading to an increased kinetic inductance $L_k$ and therefore a lower resonance frequency $f_r$. The decrease in resonance frequency $\left( f_r (|\vec{B}|) - f_r (0) \right) / f_r (0)$ is proportional to $|\vec{B}|^2$ \cite{Healey2008,Samkharadze2016,Kroll2019}. Secondly, it may create vortices in the film, which can move upon application of microwaves and cause losses and a hysteresis of the resonance frequency $f_r (|\vec{B}|)$. This can be partially mitigated using flux-trapping holes or fractal resonator geometries \cite{DeGraaf2012}. To keep $Q_i$ as high as possible, we restrict the magnetic field to the resonator plane. The single-spin control and detection experiments are not impeded by this limitation, because we can choose the substrate orientation accordingly.

To ramp up the magnetic field, we start with a low magnetic field amplitude around $\SI{0.5}{\milli\tesla}$ and rotate its orientation while monitoring the resonator frequency $f_r (\vec{B})$. The magnetic field is within the resonator plane when the resonance frequency is maximal. This procedure is repeated for increasing magnetic field strengths up to \SI{500}{\milli\tesla} because the optimal field direction can change as a function of the field amplitude $| \vec{B} |$ due to residual uncompensated magnetic fields of weakly magnetic components in the setup or the earth. We found that a few $\SI{100}{\micro\tesla}$ orthogonal to the resonator plane suffice to introduce vortices, hence the angular range within which the magnetic field is scanned is reduced as its amplitude increases.

\begin{figure}
\includegraphics{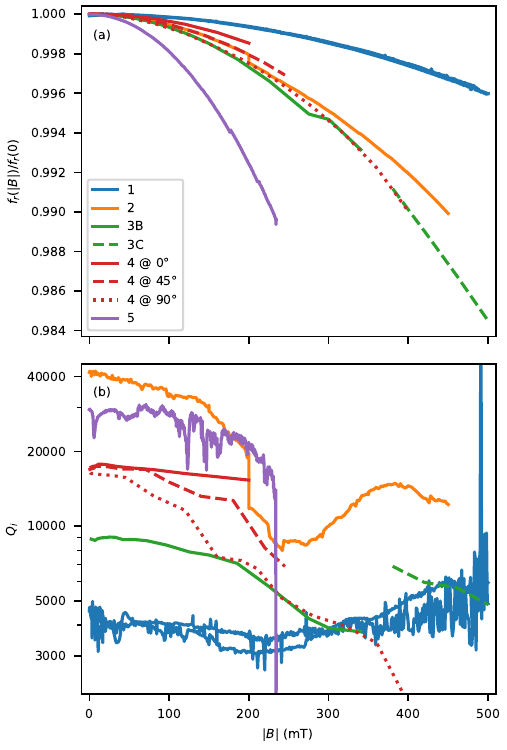}
\caption{(a) Relative change of the resonance frequency and (b) $Q_i$ as a function of the in-plane magnetic field strength $| \vec{B} |$. Sample~4 was tested in fields at \SI{0}{\degree}, \SI{45}{\degree} and \SI{90}{\degree} with respect to the nanowire.}
\label{fig:magFieldSweep}
\end{figure}

Fig.~\ref{fig:magFieldSweep}a shows that all resonators follow the predicted quadratic tuning. We would expect the resonators with the smallest nanowire width $w$ to show the strongest tuning. While sample~5 tunes more than the samples with $w = \SI{4}{\micro\meter}$, sample~1 tunes the least, despite having a $w = \SI{300}{\nano\meter}$ nanowire. We do not fully understand why sample~1 tunes the least despite having the narrowest nanowire. A possible explanation can be uncompensated out-of-plane field due to inaccurate alignment or field inhomogeneity within the sample area. Sample~3 is a chip with 3 antenna-coupled resonators, referred to as 3A, 3B and 3C (see Supplementary Information \cite{SupMat}). At $|\vec{B}| = 0$, resonator 3A was strongly hybridized with the 3D box mode. Therefore, we observed resonator 3B in the range \SIrange{0}{340}{\milli\tesla}. However, as it tuned too close to the 3D box resonance frequency, an accurate measurement of its $Q_i$ was no longer possible. Hence we switched to resonator 3C for the remaining ramp to \SI{500}{\milli\tesla}. At this magnetic field strength it was detuned by $f_r (|\vec{B}|) - f_r (0) = \SI[round-mode=places,round-precision=0]{-121.971828413557}{\mega\hertz}$ from its zero-field resonance frequency.

Fig.~\ref{fig:magFieldSweep}b shows $Q_i$ during the magnetic field ramps. The general tendency is a decrease in $Q_i$ with increasing field strength, which we attribute to trapped vortices.
 The $Q_i$ of sample~1 remains approximately constant, which matches the observation that its resonance frequency changes the least. For this sample, we recorded data during the ramp-up and -down and found only a small hysteresis, indicating that very few vortices were trapped. This emphasizes the importance of careful field alignment and good field homogeneity. In sample~2 we find a clear reduction of $Q_i$ in the range $| \vec{B} | \in [\num{200}, \num{300}]~\unit{\milli\tesla}$. We think this is due to a large ensemble of inhomogeneously broadened $g \approx 2$ spins \cite{Brodsky1969}. The sharp drop in $f_r$ and $Q_i$ at $| \vec{B} | = \SI{200}{\milli\tesla}$ is due to a vortex trapped in the resonator or ground plane during the magnetic field alignment. Sample~5 shows multiple sharp drops in $Q_i$, which might originate from various spins species being tuned in and out of resonance with the resonator. It is the most sensitive to this type of loss, because of its narrow nanowire and high $Q_i$ at zero field. In the field ramp its frequency and $Q_i$ significantly reduced when passing \SI{234}{\milli\tesla}, due to vortices.

In conclusion, the resonators are quite sensitive to magnetic field misalignment. Nevertheless, given careful alignment and good field homogeneity, it is possible to maintain $Q_i > 10^4$ at $| \vec{B} | = \SI{450}{\milli\tesla}$.

\section{Application: Single-Spin Detection} \label{sec:outlook}

In the following, we provide two quantitative examples of single-spin detection protocols. One is the currently employed detection by photon counting \cite{Wang2023,Travesedo2025,OSullivan2025}, the other is the dispersive readout, which is widely used for superconducting qubits \cite{Blais2021}.

\subsection{Photon counting}

\begin{figure}
\includegraphics{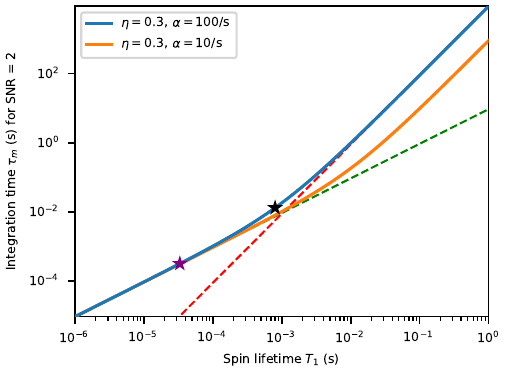}
\caption{Integration time $\tau_m$ to reach $\text{SNR} = 2$ with photon counting. The black star indicates the conditions in Ref.~\cite{OSullivan2025}, the purple star indicates the conditions expected from the resonators in this work. The red and green dashed lines mark the dark-count limited and shot-noise limited regimes, respectively.}
\label{fig:photonCountingRegimes}
\end{figure}

The increased coupling between spins and the resonator decreases the spin excited state lifetime $T_1$ and therefore allows for a faster collection of fluorescence microwave photons. If we assume one emitted photon per $T_1$, the number of photons emitted over a measurement time $\tau_m$ is $N = \tau_m / T_1$. In a detection setup with overall efficiency $\eta$, this will lead to $\eta N$ detected photons, plus $\alpha \tau_m$ dark counts, where $\alpha$ is the dark-count rate. The signal-to-noise ratio $\text{SNR}$ after an integration time $\tau_m$ is \cite{Balembois2024}
\begin{equation}
    \text{SNR}(\tau_m) = \frac{\eta \sqrt{\tau_m / T_1}}{\sqrt{\alpha \tau_m + \eta (1-\eta)}} .
\end{equation}
To reach a given $\text{SNR}$, one must integrate the signal for
\begin{equation}
    \tau_m = \left( \frac{\text{SNR}}{\eta} \right)^2 T_1 \left( \vphantom{\hat{a}} 2 T_1 \alpha + \eta (1-\eta) \right) .
\end{equation}
If the contribution of dark counts is small compared to the spin fluorescence $2 T_1 \alpha \ll \eta (1-\eta)$, the shot noise dictates the necessary measurement duration. In this regime, the measurement time to reach a given $\text{SNR}$ is $\tau_m \sim T_1$. If instead, the dark counts contribute the majority of detection events, it takes much longer to discriminate the signal from a single spin against the background and we find $\tau_m \sim {T_1}^2$. In a state-of-the-art setup \cite{OSullivan2025}, $T_1 = \SI{0.8}{\milli\second}$, $\eta = \num{0.3}$ and $\alpha = \SI{100}{\per\second}$, which is right between the two regimes, as seen in Fig.~\ref{fig:photonCountingRegimes}.

The example spin used throughout this paper, with $g_0 = 2\pi \cdot \SI{30.0}{\kilo\hertz}$, has a lifetime of $T_1 = \SI[round-mode=places,round-precision=0]{33.1572798108115}{\micro\second}$ when resonantly coupled to a resonator with $\kappa = 2\pi \cdot \SI{750}{\kilo\hertz}$. This reduction of $T_1$ by a factor of \num[round-mode=places,round-precision=0]{24.1274315795696} leads to a reduction in measurement time by a factor of \num[round-mode=places,round-precision=0]{41.20892621429833}.

\subsection{Dispersive readout}

With the increased spin-resonator coupling rate $g_0$ it may become possible to read out the state of a single spin through the shift if the resonator frequency in the dispersive regime $\Delta \gg \kappa$, where $\Delta$ is the detuning between the spin and the resonator. The quantum non-demolition nature of this readout allows one to determine the state of a single spin in a single shot without repeatedly exciting the spin. This mitigates a problem in the readout of nearby nuclear spins: Every time the electron spin decays it might flip the state of the nuclear spin \cite{OSullivan2025}. This source of nuclear-spin readout infidelity can be greatly reduced with the dispersive electron spin readout. Furthermore, the hardware required for dispersive readout is standard quantum microwave equipment as compared to a microwave single-photon detector.

\begin{figure}
\includegraphics{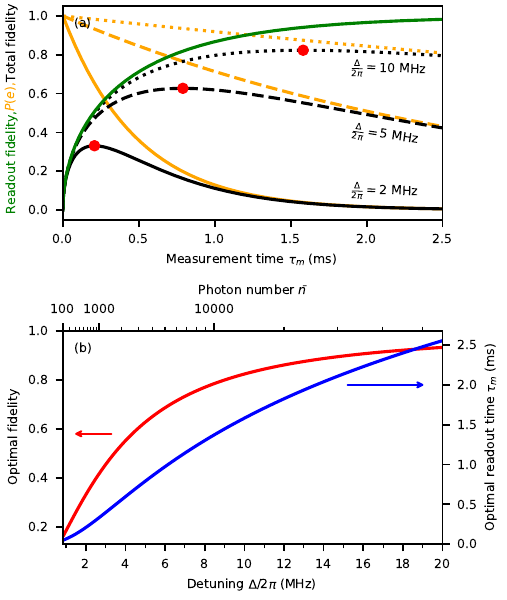}
\caption{(a) The readout fidelity $\mathcal{F}_r$, excited state probability $P(e)$ and total fidelity $\mathcal{F}$ as a function of the measurement time $\tau_m$. (b) The peak value of the fidelity $\mathcal{F}$ and the readout times $\tau_m$ at which they occur as a function of the detuning $\Delta$.}
\label{fig:dispersiveReadout}
\end{figure}

In the dispersive regime, the resonator frequency is $\omega_r \mp \chi = \omega_r \mp {g_0}^2 / \Delta$ when the spin is in the ground/excited state. When the resonator is probed at its bare frequency $\omega_r$, its reflection coefficient is
\begin{equation}
    S_{11, g/e} = \frac{\kappa_c - \kappa_i \pm 2i\chi}{\kappa_c + \kappa_i \mp 2i\chi}. \label{eq:s11}
\end{equation}
The $\text{SNR}$ to discriminate between these two reflection coefficients is
\begin{equation}
    \text{SNR} = \sqrt{2 \dot{N}_\text{in} \tau_m \eta} \left| S_{11, e} - S_{11, g} \right|,
\end{equation}
where $\eta$ is the measurement efficiency, taking into account losses and added noise, and $\dot{N}_\text{in} = P_\text{in} / (h f_r)$ is the rate of incident photons, given a drive power $P$. Here, we assume that the measurement time is much longer than the resonator lifetime $\tau_m \gg 1/\kappa$, so that the system is in a steady state throughout the measurement. Using the relation
\begin{equation}
    \bar{n} = \dot{N}_\text{in} \frac{\kappa_c}{(\kappa / 2)^2 + \chi^2},
\end{equation}
the $\text{SNR}$ becomes
\begin{equation}
    \text{SNR} = \sqrt{\frac{8 \bar{n} \tau_m \eta \kappa_c}{(\kappa / 2)^2 + \chi^2}} | \chi |
    \approx \sqrt{8 \bar{n} \tau_m \eta \kappa_c} \frac{| \chi |}{\kappa / 2}.
\end{equation}
For resonator photon numbers exceeding the critical photon number $n_\text{crit} = \Delta^2 / \left( 4 {g_0}^2 \right)$ the non-demolition character of this readout method can be violated. Hence, we limit the mean resonator photon number to $\bar{n} = n_\text{crit} / 2$ and arrive at a detuning-independent
\begin{equation}
    \text{SNR} = 2 \frac{g_0}{\kappa} \sqrt{\tau_m \eta \kappa_c}.
\end{equation}
The readout fidelity for a given $\text{SNR}$ is $\mathcal{F}_r = \text{erf} \left( \text{SNR}/2 \right)$. The measurement time $\tau_m$ is limited by the excited-state decay time $T_1$ of the spin.
\begin{equation}
    \frac{1}{T_1} = \Gamma_\mathrm{P} + \gamma = \frac{4 {g_0}^2}{\kappa} \frac{(\kappa / 2)^2}{(\kappa / 2)^2 + \Delta^2} + \gamma,
\end{equation}
where $\Gamma_\mathrm{P}$ is the Purcell-enhanced decay rate into the resonator and $\gamma$ the decay rate into all other modes, dominated by non-radiative decay. The latter is usually on the order of $10^{-3} \, \text{/s}$ for nitrogen vacancy centers \cite{Abobeih2018} and donors in silicon \cite{Bienfait2016} or $1 \, \text{/s}$ for lanthanides \cite{LeDantec2021}. While the readout fidelity asymptotically approaches $1$ as $\tau_m \to \infty$, the population of the excited state decays as $P(e) = \exp \left( - \tau_m / T_1 \right)$, as shown in Fig.~\ref{fig:dispersiveReadout}a. We are therefore interested in the total fidelity $\mathcal{F} = P(e) \mathcal{F}_r$, denoting the overall probability to correctly detect the prepared spin state. It has an optimum at a finite $\tau_m$ which depends on the detuning $\Delta$.

Fig.~\ref{fig:dispersiveReadout}b shows the relation between the detuning $\Delta$, the corresponding resonator photon number $\bar{n}$, the measurement time $\tau_m$ which maximizes the total fidelity $\mathcal{F}$ and the maximal value of $\mathcal{F}$. For these calculations, we assume $g_0 = 2\pi \cdot \SI{30.0}{\kilo\hertz}$, $\eta = \num{0.3}$ and a critically coupled resonator with $Q = \num{10000}$. We find that a single-shot dispersive readout should be possible within a few \unit{\milli\second} with a total fidelity of more than \SI{80}{\percent}, given the detuning is high enough. A practical limit to consider is the power at the first amplifier, typically a traveling-wave parametric amplifier (TWPA). TWPAs start saturating around an input power on the order of \SI{-100}{dBm}, which corresponds to the emission of a resonator with $\kappa = 2\pi \cdot \SI{750}{\kilo\hertz}$ containing $\bar{n} = \num[scientific-notation=true,round-mode=places,round-precision=1]{4270.13125721641}$ photons. For a critically coupled resonator, however, $\bar{n}$ can be significantly higher, because $|S_{11}|$ is very small near resonance (see Eq.~\ref{eq:s11}), and consequently the power reaching the TWPA remains well below saturation.

\section{Conclusion} \label{sec:conclusion}

The parallel-plate resonators presented in this work are a promising platform for microwave-only experiments with single electron spins. The highly concentrated magnetic field of the resonator mode increases the coupling to spins by a factor of $\sim 5$ compared to single-layer resonators \cite{Wang2023,Travesedo2025,OSullivan2025}. The multiple fabrication methods and mechanisms to couple to a transmission line we presented open the possibility to use these resonators together with a large variety of spins. The resonator characterization shows that overall $Q$-factors of $10^4$ are achievable and how the resonators behave under magnetic field. We furthermore estimated a speedup of photon counting experiments by almost 2~orders of magnitude, which will make weaker coupled spins more accessible. Lastly, we quantified two methods to detect spins without single-photon detector. These detection schemes require less sophisticated setups and can therefore readily be adopted by the community.

\begin{acknowledgments}
We acknowledge technical support from P.~Simon, P.-F.~Orfila and S.~Delprat, and are grateful for fruitful discussions within the Quantronics group. We acknowledge support of the R\'egion Ile-de-France through the DIM QUANTIP, from the AIDAS virtual joint laboratory, and from the France 2030 plan under the ROBUSTSUPERQ (ANR-22-PETQ-0003) grant and NISQ2LSQ (ANR-22-PETQ-0006) grant. This project has received funding from the European Union Horizon 2020 research and innovation program under the project OpenSuperQ100+, and from the European Research Council under the grant no. 101201502 (ONESPIN). Part of the research was carried out at the Jet Propulsion Laboratory, California Institute of Technology, under a contract with the National Aeronautics and Space Administration (80NM0018D0004).
\end{acknowledgments}

\bibliography{apssamp}

\end{document}


\preprint{APS/123-QED}

\title{Supplementary Information for: Superconducting Parallel-Plate Resonators for the Detection of Single Electron Spins}
\maketitle


\section{Finite-element simulation using Ansys Electromagnetic Suite}

We model the superconductors as 2D structures with an impedance boundary condition, assigning them a reactance of $X_\Box = 2\pi f_\text{target} L_{k, \Box} = \SI{9.4}{\milli\ohm/\Box}$ with a target frequency of $f_\text{target} = \SI{7.5}{\giga\hertz}$ and a kinetic inductance of $L_{k, \Box} = \SI{0.2}{\pico\henry/\Box}$. We then adjust the size of the capacitors until the eigenmode frequency is within \SI{1}{\percent} around $f_\text{target}$. We assign a mesh operation on the nanowire, requiring all elements to have edges shorter than a third of the nanowire width. The rest of the structure is meshed automatically with the 'Classic' algorithm of the software.
We calculate the amplitude of the current vacuum fluctuations $\delta I$ from the current through the nanowire $I_{\SI{1}{\joule}}$ when there is an energy of \SI{1}{\joule} in the mode by normalizing to the average magnetic energy of the mode $\tfrac{1}{4} h f_r$ in the vacuum state:
\begin{equation}
    \delta I = I_{\SI{1}{\joule}} \sqrt{\frac{\tfrac{1}{4} h f_r}{\SI{1}{\joule}}}
\end{equation}
To determine the coupling between the resonator and the box mode, as in Sec.~II~B of the main text, we sweep the resonance frequency of the resonator across the resonance of the box mode by changing the radius of the capacitor. The two eigenfrequencies of the system show an avoided crossing where the minimum frequency difference corresponds to twice the coupling rate.

\section{Experimental setup}

\begin{figure*}
    \centering
    \includegraphics{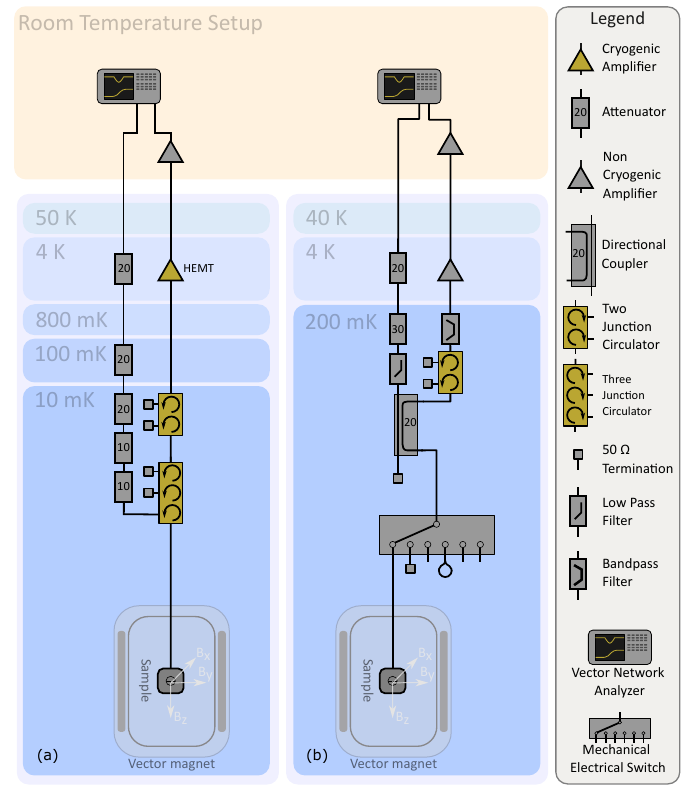}
    \caption{Experimental setup for two different cryostats, (a) dilution refrigerator setup (b)  ADR cryostat}
    \label{fig:exp_setup}
\end{figure*}

Resonator characterization was performed in two cryostats, one dilution cryostat and one Adiabatic Demagnetization Refrigeration (ADR) cryostat. Comparable setups are implemented in both cryostats, which can be seen in Fig.~\ref{fig:exp_setup}. Both setups include heavy attenuation on the input lines to enable measurements at the single-photon level. The attenuators were placed on different temperature stages to avoid the dissipation of power close to the sample stage. In the case of (a), the dilution refrigerator, one triple junction circulator is used to allow the reflection measurement and to isolate the resonator from any noise from the amplifier. The two-junction circulator in series has the same purpose. In setup (b), a directional coupler fills the role of the circulator, together with a band-pass filter to reject out-of-band amplifier noise. The mechanical switch in this setup allows for recording frequency spectra, which will be used in the de-embedding procedure described in Section \ref{sec:de-embedding}. The sample is placed in a $(1, 1, 6) \, \unit{\tesla}$ vector magnet in setup (a) and a $(0.2, 0.2, 1) \, \unit{\tesla}$ vector magnet in setup (b).

\section{De-embedding} \label{sec:de-embedding}

In the cryostat ADR setup depicted in Fig ~\ref{fig:exp_setup}b, instead of a circulator, a directional coupler is used to separate the signal reflected off the resonator from the input. Although the coupler has a higher bandwidth than a typical circulator, it causes slight impedance mismatches in the line, leading to partial reflections interfering with the resonator reflection $S_{11}(f)$. We have therefore implemented a de-embedding procedure to reconstruct the resonator reflection spectrum without the reflections.

As shown in Fig.~\ref{fig:exp_setup}b, the mechanical switch below the coupler allows us to record comparison spectra, as connections to a \SI{50}{\ohm} termination, a short, and an open end are available.

\begin{figure}
    \centering
    \includegraphics{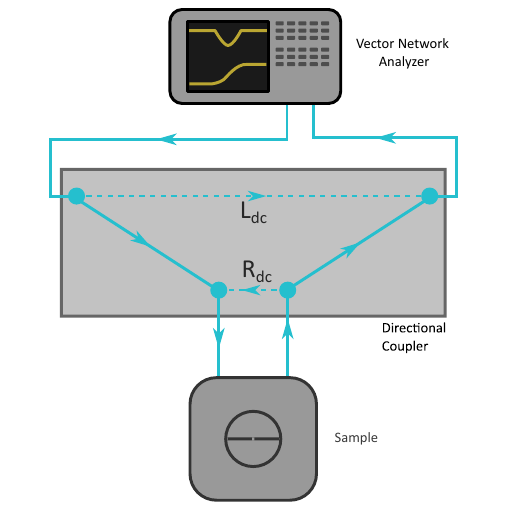}
    \caption{Signal flow diagram of the loops considered in the de-embedding procedure}
    \label{fig:deembedding_drawing}
\end{figure}

The focus of the procedure is on the directional coupler that the signal passes through before and after being reflected off the sample. We take into account (cf. Fig~\ref{fig:deembedding_drawing}) the part of the signal $L_{dc}$ which leaks through the directional coupler without interacting with the sample as well as the signal loop due to reflections off the directional coupler $R_{dc}$ after the sample. The resulting transfer function describing the  spectrum $S_{21}(f)$ measured by the VNA as a function of the reflection coefficient of the sample $S_{11}(f)$ is
\begin{equation} \label{eq:embedding}
	S_{21} = \frac{A+B S_{11}}{1+C S_{11}}
\end{equation}
where $A$, $B$, and $C$ are complex frequency-dependent parameters determined by coupling, leakage, and transmission paths. They can be eliminated by considering the three alternative terminations. For a matched \SI{50}{\ohm} termination, $S_{11, \SI{50}{\ohm}} = 0$. Ideal open and short ends return $S_{11,\text{op}} = e^{i\phi_\text{op}}$ and $S_{11,\text{sh}} = -e^{i\phi_\text{sh}}$, where the $\phi_i = 2\pi f \tau_i$ denote the accumulated phase due to different electrical delays $\tau_i$ in the different lines. Inserting these coefficients into Eq.~\ref{eq:embedding} results in 3 equations
\begin{subequations}
    \begin{align}
        S_{21, \SI{50}{\ohm}} &= A \\
        S_{21, \text{op}} &= \frac{A + B e^{i\phi_\text{op}}}{1 + C e^{i\phi_\text{op}}} \\
        S_{21, \text{sh}} &= \frac{A - B e^{i\phi_\text{sh}}}{1 - C e^{i\phi_\text{sh}}},
    \end{align}
\end{subequations}
which allow us to eliminate $A$, $B$ and $C$ in Eq.~\ref{eq:embedding} and solve for the reflection spectrum of the resonator
\begin{widetext}
\begin{equation}
S_{11} =
\frac{(S_{21, \text{op}} - S_{21, \text{sh}})(S_{21} - S_{21, \SI{50}{\ohm}})}
{
e^{i\phi_\text{sh}}(S_{21, \text{op}} - S_{21})(S_{21, \text{sh}} - S_{21, \SI{50}{\ohm}})
- e^{i\phi_\text{op}}(S_{21} - S_{21, \text{sh}})(S_{21, \text{op}} - S_{21, \SI{50}{\ohm}})
}.
\label{eq:final_S11}
\end{equation}
\end{widetext}

\begin{figure*}
    \centering
    \includegraphics{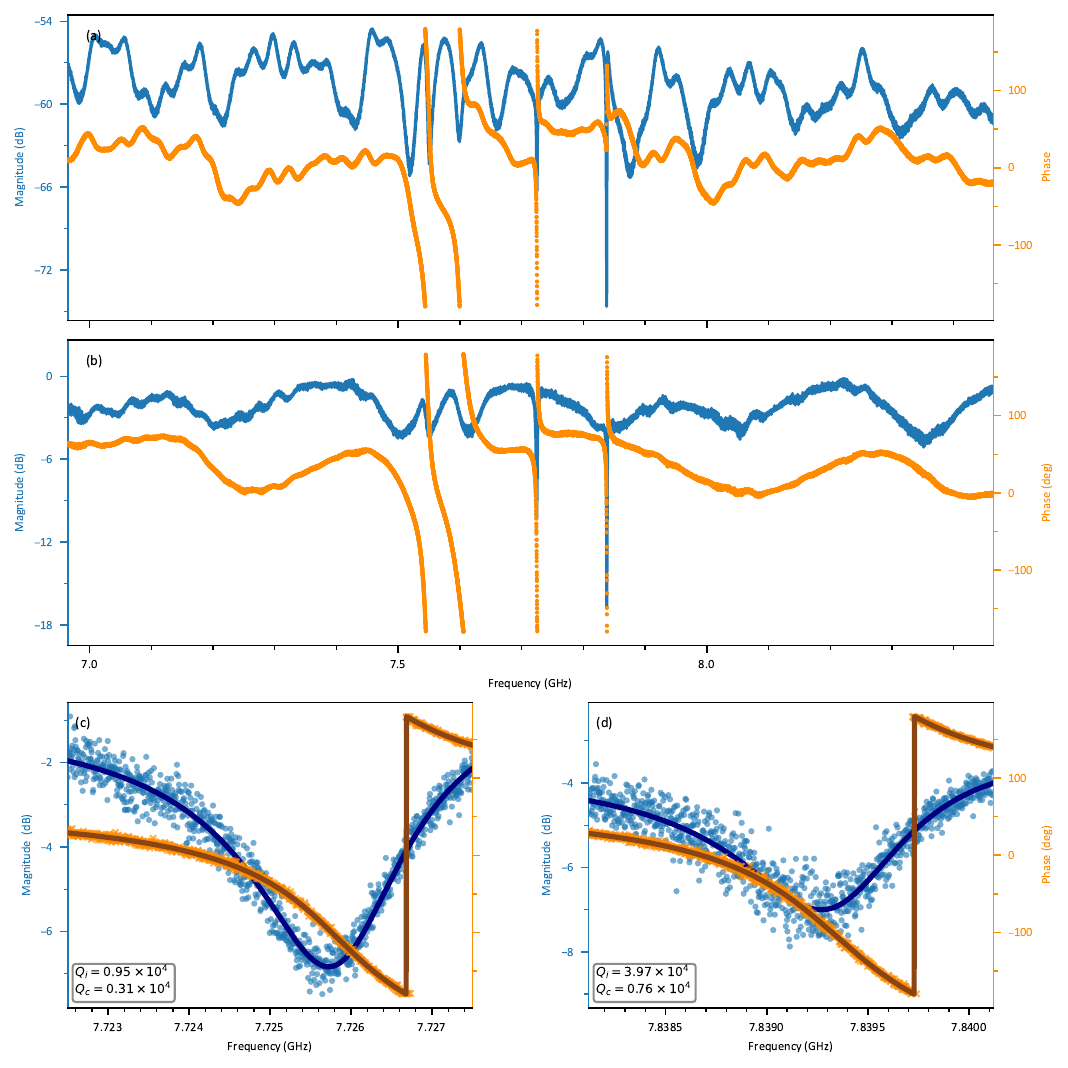}
    \caption{VNA Spectra illustrating de-embedding procedure: The broad phase and magnitude spectrum displaying 3 resonances of sample~3 and the 3D cavity mode before (a) and after de-embedding (b). (c) Resonance of resonator~3B, after de-embedding, with applied fit. (d) Resonance of resonator~3C, after de-embedding, with fit.}
    \label{fig:kiutra_deembedding}
\end{figure*}

An example application can be seen in Fig.~\ref{fig:kiutra_deembedding}. It shows the spectrum of sample~3, a chip with 3 antenna-coupled resonators. The spectrum before de-embedding (a) shows severe interference ripples in the baseline. After de-embedding, most of the spectrally narrow ripples are removed, allowing for a more accurate extraction of $Q_i$ and $Q_c$ from the fits. Diagrams (c) and (d) show fitted spectra of the two rightmost resonances. These were recorded after a thermocycle to remove magnetic vortices, which were still present in (a, b). The two broader resonances are hybridized modes of the lowest frequency resonator and the 3D cavity.

\section{Microscope images of samples}

\begin{figure*}
\includegraphics{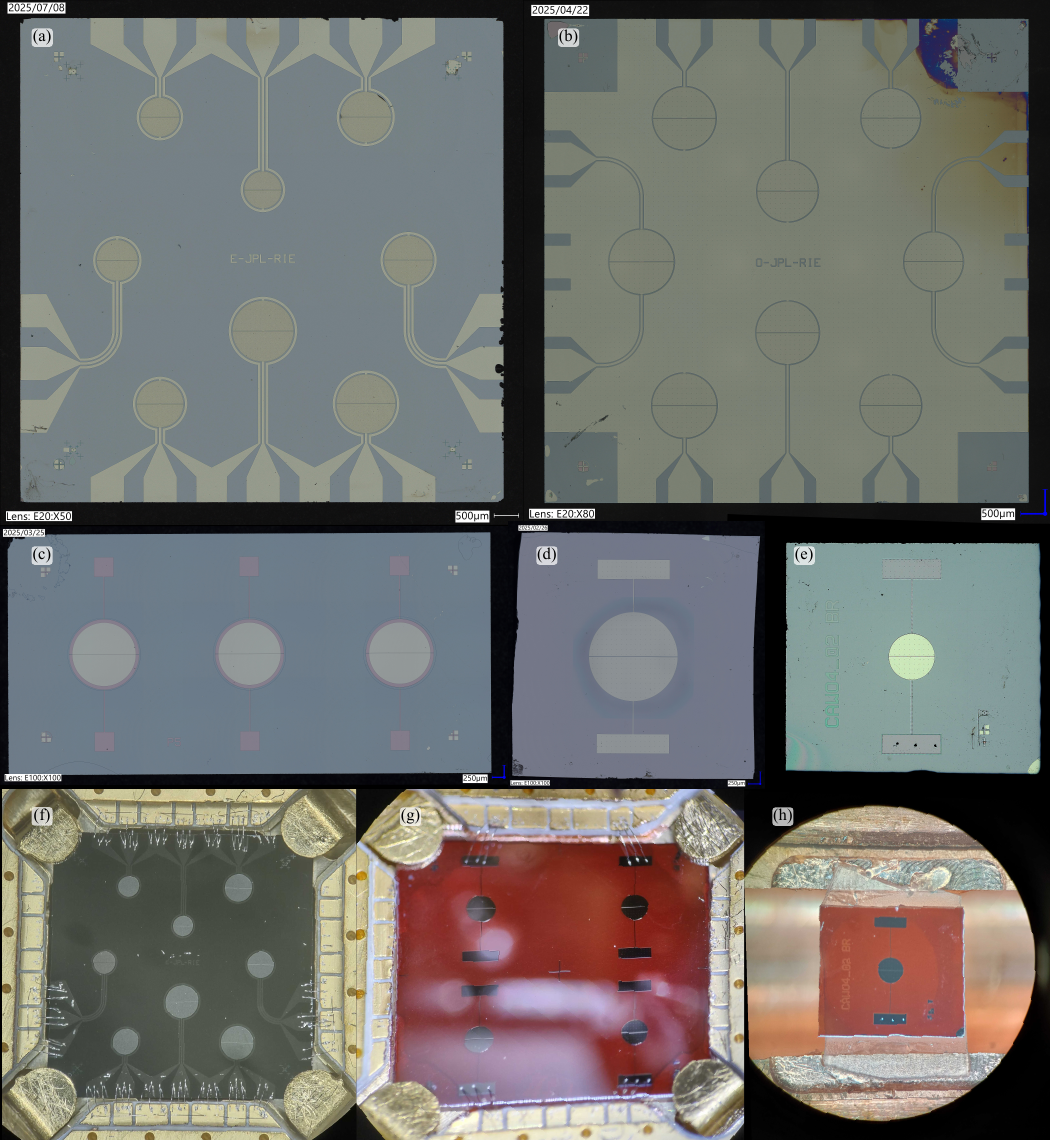}
\caption{Microscope images of the samples: (a) Sample~1, fabricated from a Nb-Si-Nb trilayer, waveguide coupled, without ground plane, \SI{4}{\micro\meter} constriction in the nanowire layer. (b) Sample~2, fabricated from a Nb-Si-Nb trilayer, waveguide coupled, with a ground plane, \SI{300}{\nano\meter} nanowire. (c) Sample~3, fabricated layer by layer, 3 antenna coupled resonators, enumerated 3A, 3B and 3C from left to right. In the slightly red region around the round capacitor the top Nb layer (counter-electrode layer) has been removed after an initial test to tune the resonators to higher frequencies. (d) Sample~4, fabricated on a membrane. The suspended part is the slightly darker square in the middle of the chip. (e) Sample~5, fabricated layer by layer on a CaWO$_4$ chip. (f) Sample~1 wirebonded. The resonator characterized in the main text is the one connected in the middle on the bottom edge. (g) Sample~5 with wirebonds to a Nb bonding pad on top of the antenna pads. This configuration was used to initially characterize the resonance frequencies of the resonators. (h) The bottom right resonator after removing the wirebonds and the bonding pad. It is attached to a piece of sapphire, which is attached to the bottom half of a 3D cavity, using vacuum grease.}
\label{fig:sample_pictures}
\end{figure*}

Microscope images of all samples listed in Table~1 of the main text are compiled in Fig.~\ref{fig:sample_pictures}.

\bibliography{apssamp}